# Magnetic Josephson Junctions with Superconducting Interlayer for Cryogenic Memory

Igor V. Vernik, Vitaly V. Bol'ginov, Sergey V. Bakurskiy, Alexander A. Golubov, Mikhail Yu. Kupriyanov, Valery V. Ryazanov and Oleg A. Mukhanov

*Abstract*— We investigate Magnetic Josephson Junction (MJJ) – a superconducting device with ferromagnetic barrier for a scalable high-density cryogenic memory compatible with energy-efficient single flux quantum (SFQ) circuits. The superconductor-insulator-superconductor-ferromagnet-superconductor (SIS'FS) MJJs are analyzed both experimentally and theoretically. We found that the properties of SIS'FS junctions fall into two distinct classes based on the thickness of S' layer. We fabricate Nb-Al/AlOx-Nb-PdFe–Nb SIS'FS MJJs using a co-processing approach with a combination of HYPRES and ISSP fabrication processes. The resultant SIS'FS structure with thin superconducting S'-layer is substantially affected by the ferromagnetic layer as a whole. We fabricate these type of junctions to reach the device compatibility with conventional SIS junctions used for superconducting SFQ electronics to ensure a seamless integration of MJJ-based circuits and SIS JJ-based ultra-fast digital SFQ circuits. We report experimental results for MJJs, demonstrating their applicability for superconducting memory and digital circuits. These MJJs exhibit $I_cR_n$ product only ~30% lower than that of conventional SIS junctions co-produced in the same fabrication. Analytical calculations for these SIS'FS structures are in a good agreement with the experiment. We discuss application of MJJ devices for memory and programmable logic circuits.

## I. INTRODUCTION

RAPID Single Flux Quantum (RSFQ) technology [1] offers unmatched clock speed and low power dissipation [2]-[4]. Recently, its power dissipation was further reduced with an introduction of energy-efficient single flux quantum logics [5]-[8], making these circuits viable contenders for the implementation of next generation high performance computing (HPC) [9], cryogenic sensor image processors, and high-end instrumentation. However to date, the high performance potential of RSFQ technology was realized only for limited practical applications of mixed signal Digital-RF circuits for communications and signal intelligence [10]-[12]. Highly promising applications in HPC, instrumentation and other fields remain unfulfilled. The chief reason for this is a lack of dense, high capacity superconducting memory capable of matching unparallel clock speed and low power available with single flux quantum technology.

In spite of substantial efforts to develop a suitable superconducting memory technology [13]-[21], only a 4 kbit random access memory (RAM) was demonstrated to date [16]. This capacity was not sufficient for practical applications. Serial memories applicable to some instrumentation applications [22] did not go further either, as a 1 kbit memory occupied a substantial area [23] making it hardly extendible to application requirements. Superconducting memory exhibits relatively low density due to relatively large size of SQUID-based memory cells coupled to address lines via transformers which are difficult to scale [13]-[21]. Additionally, the RAM ac power was identified as an obstacle for the development of larger size memories [21].

In some cases, the lack of high capacity superconducting memory can be circumvented by architectural approaches reducing memory requirements [24]. An attempt to eliminate the need for massive addressable memories by the distribution of memory functions in digital RSFQ circuits is described in [25]. However an existence of dense energy-efficient random access memory would unlock many presently not realized applications and would substantially enlarge the applicability of superconducting technologies.

In order to get around the low capacity of superconductor memories, hybrid superconductor-semiconductor schemes were pursued combining room-temperature CMOS memory and RSFQ interface circuits [26] or building a cryogenic (4 K) hybrid Josephson junction (JJ) CMOS RAM [27]. Recently, a cryogenic 64 kbit hybrid JJ-CMOS RAM was successfully demonstrated [28]. This hybrid RAM can fill a niche in the memory hierarchy, in which access time requirements are less demanding. Still, there is a strong need for a fast, energy efficient memory integrated with energy-efficient digital circuits on the same chip and capable of operating with a clock rate similar to RSFQ digital circuits.

The ideas of combining superconducting and ferromagnetic materials to build a high density cryogenic RAM were first expressed in the 90s and recently gained even more attention [29]-[33]. These ideas were based on various structures of superconducting and ferromagnetic layers, ferromagnetic dots, combined with superconducting-isolator-superconducting (SIS) JJs for readout. However, these initial concepts were not further developed. A hybrid superconducting-magnetic RAM was proposed by combining proven magnetic tunnel junctions (MTJs) with the SIS readout JJs to form a cryogenic MRAM [34].

The experimental realization of superconducting-

I. V. Vernik and O. A. Mukhanov are with HYPRES, Inc. 175 Clearbrook Rd., Elmsford, NY 10523 USA (phone: 914-592-1190; fax: 914-347-2239; e-mail: vernik@hypres.com).
V. V. Bol'ginov is with Institute of Solid State Physics, Russian Academy of Sciences, Chernogolovka, 142432 Russia (e-mail: bolg@issp.ac.ru).
S. V. Bakurskiy and M. Yu. Kupriyanov are with Skobeltsyn Institute of Nuclear Physics, Moscow State University, Moscow, 119992 Russia (e-mail: mkupr@pn.sinp.msu.ru).
A. A. Golubov is with the Faculty of Science and Technology and MESA+ Institute for Nanotechnology, University of Twente, 7500 AE Enschede, The Netherlands (e-mail: A.A.Golubov@tnw.utwente.nl).
V. V. Ryazanov is with Institute of Solid State Physics, Russian Academy of Sciences, Chernogolovka, 142432 Russia and InQubit, Inc., 21143 Hawthorne Blvd., Torrance, CA 90503, USA (e-mail: ryazanov@issp.ac.ru).

ferromagnetic junctions in 1999 [35], [36] stimulated memory ideas based on using magnetic Josephson junctions (MJJs), in which ferromagnetic layers are integrated within a Josephson junction [37]-[46]. The MJJ critical current can change and retain its value by the ferromagnetic layer magnetization allowing the realization of two distinct states with high and low $I_c$, corresponding to logical "0" and "1" states, respectively. In [44], the storage MJJs are suggested to be integrated with readout SIS JJs into SQUID loops following known superconducting RAM architectures, e.g., [14]. However, it is not evident if MJJs in the proposed configuration can bring any advantage in density and energy-efficiency as compared to the design with proven MTJs, cf. [34].

In contrast, the MJJs are suggested to perform both data storage and readout functions in [38]-[40], [45], [46]. The first MJJ-based memory with energy-efficient SFQ switching readout was proposed in [39], [45]. Such a memory required MJJs to be compatible in speed and critical currents with conventional SIS junctions. Memory built with these MJJs is electrically and physically compatible with SFQ-type circuits allowing a fabrication of memory and digital circuits on the same chip in a single process expandable to a high density 3D integration. This will enlarge the applications of MJJs beyond just memory arrays into programmable logic and enable the smaller footprint as SQUID loops are no longer required.

Our MJJs are made using only one ferromagnetic layer (F-layer). By applying magnetic field pulses (e.g., by current pulses through a superconducting Write line), the junction F-layer can be magnetized in two opposite directions. To discriminate these directions, a read current bias is applied through the MJJ inducing a reference magnetic field (self-field). Depending on F-layer magnetization, this field either adds to or subtracts from F-layer magnetic field effectively forming two possible magnetic states with high or low magnetizations corresponding to low ("1") and high ("0") MJJ critical currents, respectively (see Fig. 2 in [46] for clarity).

The switching between "0" and "1" states has been observed in Superconductor-Ferromagnetic-Superconductor (SFS) Nb/Pd$_{0.99}$Fe$_{0.01}$/Nb junction [38]. The characteristic voltage $I_cR_n$ of these SFS devices was 2-3 nV, which makes them too slow (~MHz rate) for the memory designs discussed in [39]. Achieving MJJ switching speed comparable to that of conventional JJs is essential for both types of junctions to be integrated into a single circuit operating in an SFQ non-hysteretic switching regime. By inserting an additional isolation tunnel layer (I-layer) in the junction (i.e., fabricating an SIFS structure), we were able to increase $V_c = I_cR_n$ to ~1 mV achieving high switching frequency. The SIFS MJJ based on Nb-Al/AlO$_x$-Pd$_{0.99}$Fe$_{0.01}$-Nb tunnel junctions with $V_c$ from 100 to 400 μV were reported in [39]. Very recently, we demonstrated Nb-Al/AlO$_x$-Nb-Pd$_{0.99}$Fe$_{0.01}$-Nb MJJ device of SIS'FS type where superconductivity of S'-layer was substantially affected by the adjacent ferromagnet making it close to the SIFS configuration [46]. By applying magnetic field pulses, the MJJ was switched between "0" and "1" logic states. This MJJ exhibited high characteristic voltage $I_cR_n$ ~1 mV enabling fast non-destructive readout and making it applicable for an energy-efficient memory compatible to SFQ circuits.

However, the fabricated SIS'FS showed significant differences in observed critical currents. From our initial measurements [46], it was difficult to establish whether superconductivity in the S'-layer was completely or only partially suppressed.

In this paper, we theoretically and experimentally investigate SIS'FS junctions focusing on effects of the S'-layer thickness. Our study is guided by requirements of these devices for memory applications and their compatibility to SIS-based digital circuits.

## II. SUPERCONDUCTOR-INSULARTOR-SUPERCONDUCTOR-FERROMAGNET-SUPERCONDUCTOR JUNCTIONS

### A. Theory

Here, to analyze SIS'FS devices, the conditions of dirty limit are presumed to be valid for all superconducting and ferromagnetic layers. Additionally, the effective electron-phonon coupling constant for the F is set to zero. With these assumptions, the critical current $I_c$ is calculated by numerically solving Usadel equations [47] with Kupriyanov-Lukichev boundary conditions [48]. Figures 1-3 summarize the results of this self-consistent approach.

Fig. 1 presents $I_cR_n$ product as a function of F layer exchange energy, $h$, calculated at $T=0.5T_C$ for different thickness, $L_{S'}$, of S' film, large thickness of S electrodes and F layer thickness, $L_F=2\xi_F$ with $\xi_F$ being decay length in ferromagnet, of a SIS'FS sandwich. For simplicity, in our calculations the critical temperature, $T_C$, for S and S' layers is assumed to be the same with decay lengths $\xi_S$ and $\xi_{S'}$ are, also, the same for all the films. All calculations are performed for SIS'FS structure with following layer thicknesses: for base S electrode of $2\xi_S$, for ferromagnetic layer of $2\xi_F$ and counter S electrode of $10\xi_S$. Both S'F and FS interfaces can be described by suppression parameters $\gamma=(\rho_S\xi_S)/(\rho_F\xi_F)=1$, $\gamma_B=(R_{S'F}A)/(\rho_F\xi_F)=0.3$, while suppression parameter at SI interface, $\gamma_{BI}=(R_NA)/(\rho_S\xi_S)=1000$. Here $\rho_S$ and $\rho_F$ are resistivity of S, S' and F films, respectively, $R_N$ and $R_{S'F}$ are the resistances of SI and S'F interfaces; and $A$ is the area of junction cross section.

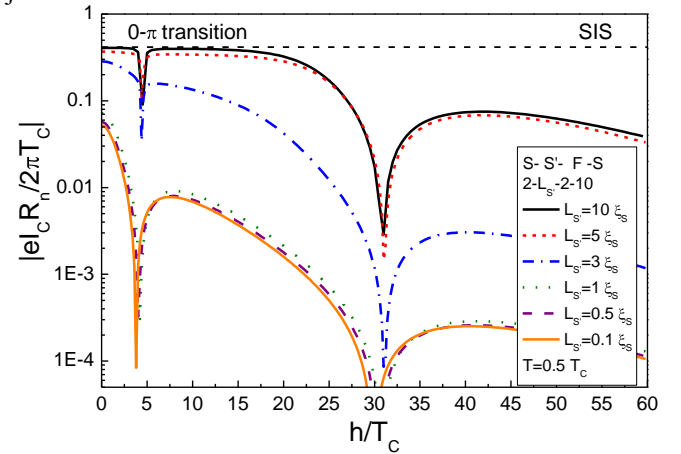

Fig. 1. $I_cR_n$ product of SIS'FS structure as a function of exchange energy $h$ of F layer calculated for a set of thickness of S' layer $L_{S'}$ and $T = 0.5\ T_C$, $L_F=2\xi_F$, $\gamma=1$, $\gamma_B= 0.3$, $\gamma_{BI}= 1000$. Superconducting and ferromagnetic layer thicknesses are normalized to $\xi_S$ and $\xi_F$, respectfully.

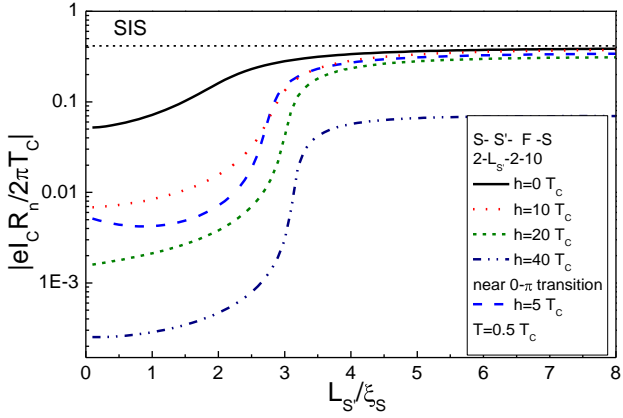

Fig. 2. $I_cR_n$ product of SIS'FS structure as a function of thickness of S' layer $L_{S'}$ calculated for a set of exchange energy $h$ and $T = 0.5T_C$, $L_F=2\xi_F$, $\gamma=1$, $\gamma_B= 0.3$, $\gamma_{BI}= 1000$. Superconducting and ferromagnetic layer thicknesses are normalized to $\xi_S$ and $\xi_F$, respectfully.

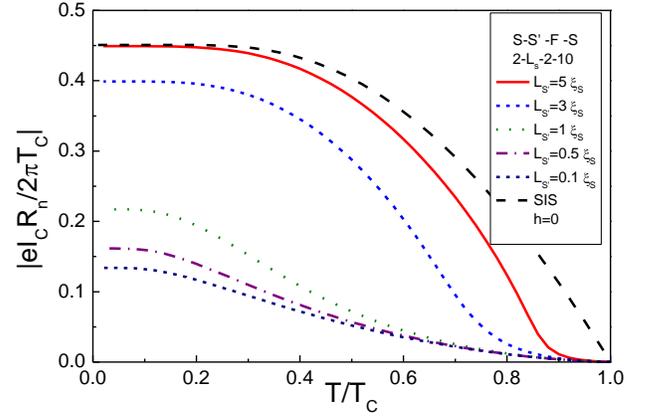

Fig. 3. $I_cR_n$ product of SIS'FS structure as a function of temperature, $T$, calculated for a set of thickness of S' layer and $h=0$, $L_F=2\xi_F$, $\gamma=1$, $\gamma_B= 0.3$, $\gamma_{BI}= 1000$. Superconducting and ferromagnetic layer thicknesses are normalized to $\xi_S$ and $\xi_F$, respectfully.

In the limit of large $L_{S'}$, the SIS'FS device can be considered as a composition of two independent SIS' and S'FS junctions coupled in series. A flowing across SIS'FS supercurrent

$$I(\varphi) = \frac{I_{SIS'} I_{S'FS} \sin(\varphi)}{\sqrt{I_{SIS'}^2 + I_{S'FS}^2 + 2I_{SIS'} I_{S'FS} \cos(\varphi)}}, \quad (1)$$

depends on the critical currents, $I_{SIS'}$ and $I_{S'FS}$, of these junctions and superconducting electrode order parameter phase difference, $\varphi$ [49].

For the given set of parameters, $L_S=10\xi_S$ and $h=0$, $I_{S'FS}$ appears to be much larger compared to $I_{SIS'}$ and in accordance with Eq. (1) the $I(\varphi) \approx I_{SIS'}\sin\varphi$. Near $h=0$, the magnitude of $I_{SIS'}$ coincides with prediction of Ambegaokar-Baratoff (AB) theory (horizontal dashed line in Fig. 1). Further increase of $h$ leads to suppression of $I_{S'FS}$. In the vicinity of $4T_C$ the S'FS junction exhibits first $0 - \pi$ transition (marked in Fig. 1) with its critical current drastically decreasing, crosses zero and increases again to its previous value. Fig. 1 presents absolute value of $I_c$ with negative values being reflected to positive. Following Eq. (1), this is accompanied by transformation of $I(\varphi)$ relationship and final generation of the $\pi$-shift between phases of order parameters of S' and S films with further $h$ increase. Note, that in the range of $h$ between $5T_C$ and $20T_C$, the SIS'FS device is $\pi$-junction with $I_cR_n$ product practically equal to that of SIS tunnel junction.

In a vicinity of $h \approx 30T_C$ there is $\pi - 0$ transition and for larger $h$ the junction is in the $0$-state with the weak place located in the F layer ($I_{S'FS} < I_{SIS'}$).

The decrease of $L_{S'}$ causes the suppression of superconductivity in the S' film. It is still weak for $L_S=5\xi_S$. With further decrease, for $L_{s'} = 3.5\xi_S$, the $I_cR_n$ product of the structure even at $h=0$ is about 0.7 normalized to the AB value. In the $\pi$ – state the $I_cR_n$ is further suppressed with $h$ increase and at $h > 10 T_C$ the structure behaves like SINFS junction. Note, that for $L_{S'} < \xi_S$, we are in the SINFS regime for any values of $h$.

Fig. 2 presents $I_cR_n$ product as a function of S' layer thickness, $L_{S'}$, calculated at $T=0.5T_C$ for different exchange energy, $h$, infinitely large thickness of S electrodes and F film thickness, $L_F=2\xi_F$.

It is seen that at $h < 20T_C$ and $L_{S'} >4\xi_S$ the magnitude of the junction critical current $I_c = I_{SIS'}$ closes to that of Ambegaokar-Baratoff (AB) theory (horizontal dashed line in Fig. 2), while the structure is either in 0- ($h < 5T_C$) or in the $\pi$-state ($5T_C < h < 30T_C$) depending on the value of exchange energy, $h$ (see Fig. 1). At lower thicknesses, the superconductivity in the layer S' is suppressed due to the proximity effect.

Fig. 3 presents the temperature dependencies of $I_cR_n$ product calculated for different thickness, $L_{S'}$, of S' film, infinitely large thickness of S electrodes and F film thickness, $L_F=2\xi_F$, $\gamma=1$, $\gamma_B=0.3$, $\gamma_{BI}=1000$ and $h=0$. In a vicinity of $T_c$ all the curves have a positive curvature. This behaviour is typical for SINS structures. It comes from suppression of superconductivity at SN interface and in the S' layer, which is the stronger the closer is the temperature to $T_c$. For smaller $T$ the intrinsic superconductivity in S' film and at FS interface are restored. At $L_{S'} > 3\xi_S$ this is accompanied by strong increase of $I_c$ to value, which is close to that followed from AB theory for SIS tunnel junctions (dashed line in Fig. 3). For $L_{S'} = \xi_S$ the shape of $I_c(T)$ is typical for SINS structures. It is proportional to $(T_c - T)^{3/2}$ at $T = T_c$ and saturates for $T = E_{Th}$, where $E_{Th}$ is Thouless energy [50].

The larger is $h$, the larger is the thickness of $L_{S'}$ at which the crossover between this two typical shapes of $I_c(T)$ occurs.

In the interval of S' layer thickness $3\xi_S < L_{S'} <5\xi_S$, the SIS'FS structure, when it is biased, behaves as a composition of two independent SIS' and S'FS junctions coupled in series. Contrary to that, when it is placed in external magnetic field, $H_{ext}$, it behaves a single junction. The S' layer is too thin to screen $H_{ext}$.

### B. Fabrication

Our fabrication process is based on a co-fabrication approach using a combination of HYPRES and Institute of Solid State Physics (ISSP) fabrication processes. Below we briefly trace our steps to produce samples. The details of our co-fabrication, i.e., techniques and methods used, thicknesses of layers, etc., reported elsewhere [46]. We produced a series of wafers with an in-situ deposited Nb-Al/AlO$_x$-Nb trilayer targeted for 4.5 kA/cm$^2$ Josephson critical current density standard for HYPRES technology [51]. Next, the wafers were diced into samples and sent to ISSP, where after about 10 nm of Nb top electrode was etched away the PdFe/Nb bilayer was deposited. The ferromagnetic layer was thin enough to avoid significant critical current suppression. Then, we formed a square mesa of 10 x 10 μm² sizes and patterned the bottom Nb-electrode. At the next step, we evaporated an isolation layer with a contact window with 4 x 4 μm² junction contact size. Last, the Nb wiring electrode was deposited and patterned.

Depending on S' thickness, the resultant SIS'FS structure can be approximated by a single SIFS junction or by two SIS' and S'FS junctions. The former case is expected when superconductivity of S'-layer is substantially suppressed by the adjacent ferromagnet resulting in SIFS configuration. For a reference, we also produced conventional SIS'S junctions on the samples from the same wafers by excluding the F-layer deposition step.

### C. Experimental Results

Measurements were performed in a variable temperature liquid He cryostat with samples placed in a vacuum can with He gas added for better heat exchange. Fig. 4 shows the measured current-voltage (*I-V*) characteristics at 4.2 K of SIS'FS and SIS'S junctions produced in the similar fabrication cycles with only a positive quadrant shown for simplicity. The *I-V* curve of the reference SIS'S junction shows critical current $I_c = 3.3$ mA. The observed $I_c$ corresponds to critical current density consistent with the target of 4.5 kA/cm² and the reduced gap in Nb counter electrode. The *I-V* curve shows smaller value of sum of superconducting Nb electrode gaps than 2.6 mV gap voltage observed for JJs made by the standard HYPRES process [51]. We attribute this to a gap suppression in the thin Nb counter electrode [46]. Both SIS'S and SIS'FS *I-V* curves exhibit a step at ~ 700 μV that can be explained by the peculiarity at the gap difference due to different quality of the base and counter Nb electrodes. The negative slope of the *I-V* curve may be caused by the uniform change in the energy gap due to influence of quasiparticle injection [52] resulting in a non-equilibrium state of junctions [53, 54].

This SIS'FS junction at $T = 4.2$ K exhibits characteristic voltage $I_cR_n$ of ~700 μV making it fairly compatible to the speed of conventional SIS JJs used in RSFQ circuits.

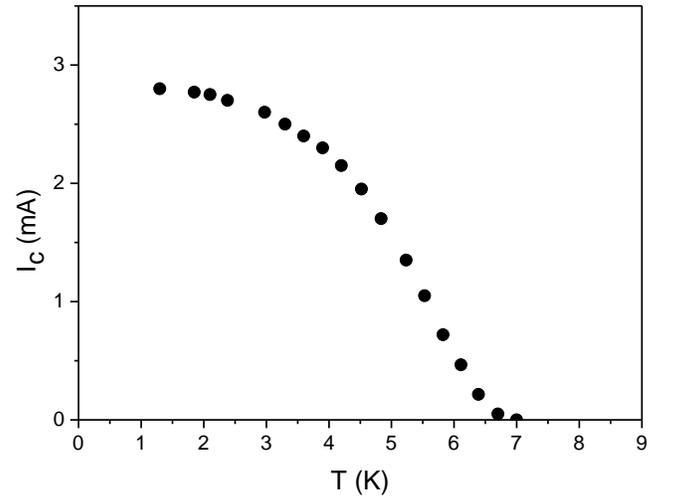

Fig. 5. $I_c(T)$ dependence for SIS'FS junction with F-layer thickness of 14 nm.

Fig. 5 shows $I_c(T)$ dependence for SIS'FS junction with F-layer of 14 nm. The *I-V* curve for this junction at $T = 4.2$ K is presented in Fig. 4. $I_c$ significantly varies from 3 K to 7 K temperature range reaching its maximum at lower temperatures. The dependence mimics the calculated curves shown in Fig. 3 for $L_{S'}=3$-$5\xi_S$. This may indicate transition of residual S' layer into superconducting state at temperature low than 4.5-5 K. A comparison of the experimental and calculated curves is only semi-quantitative due to simplicity of our model. To be sure that S'-interlayer is superconductive, we measured $I_c$ dependence from external magnetic field $H_{ext}$ at this lower temperature of 1.3 K.

Fig. 6 demonstrates a typical Fraunhofer-like $I_c(H_{ext})$ dependence of SIS'FS critical current on external magnetic field $H_{ext}$ aligned parallel to the layers within the junction and supplied by an external solenoid. The external field is swept from 100 Oe, well above saturation filed for our $Pd_{0.99}Fe_{0.01}$ layer, to -100 Oe and back. The $I_c(H_{ext})$ dependence with $H_{ext}$ decreasing is shown by solid circles and dotted line and $I_c(H_{ext})$ with $H_{ext}$ increasing is shown by open circles and dashed line. For each measured point of the $I_c(H_{ext})$ curve, the current through the sample is swept from a subcritical value upwards until the threshold voltage is exceeded while the external magnetic field is fixed. The measurement is performed for the SIS'FS that was externally post-process shunted by a small ~ 0.02 Ω resistor made of Al wire. With this shunting SIS'FS does not require the reduction of the bias current to almost zero for each $I_c(H_{ext})$ measurement point. In addition, the shunted SIS'FS is much less susceptible to the external noise.

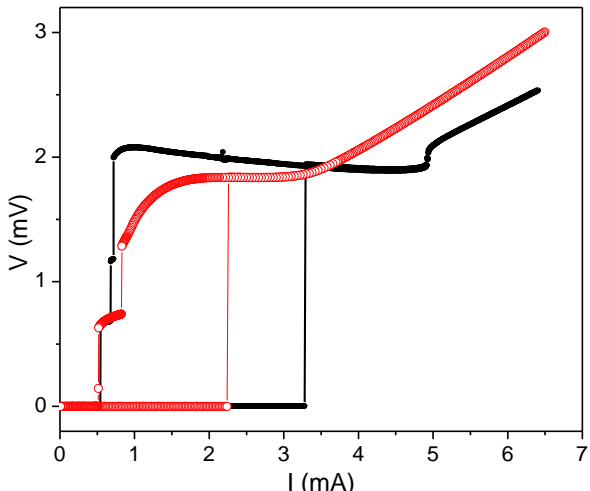

Fig. 4. Current-voltage characteristics at 4.2 K of SIS'FS (open circles) and reference SIS'S (closed circles) with 14-nm ferromagnetic layer Josephson junctions fabricated on the same wafer. Data from [46] replotted here.

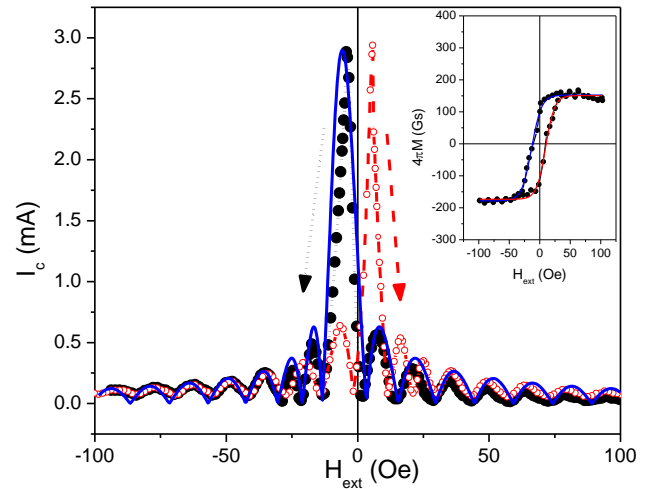

Fig. 6. Hysteretic dependence of critical current $I_c$ on external magnetic field $H_{ext}$ for SIS'FS junction measured at T=1.3 K. Magnetic field sweep directions are shown by arrows. Inset shows magnetization curve *M(H)*.

The application of an external magnetic field changes the magnetization of the ferromagnetic layer that in turn changes the junction $I_c$ when junction bias current is applied for readout. This realizes two distinct states with high and low $I_c$, corresponding to logical "0" and "1" states, respectively. SIS'FS junction switching curves similar to one reported in [46] were observed when junction switched from a superconducting (logical "0") to a resistive state (logical "1") and back by a pulse of magnetic field. This switching curves were recorded at $T$ =1.3 K when junction was biased with 1.5 mA for readout.

We also performed Josephson magnetometry [38] for SIS'FS junction with 14 nm thick F layer using well-known feature, that magnetic flux through Josephson junction equal to natural multiples of $\Phi_0$ ($\Phi_0$, $2\Phi_0$, $3\Phi_0$ …) in minima of $I_c(H_{ext})$ curve and odd integer half of $\Phi_0$ ($1/2\Phi_0$, $3/2\Phi_0$, $5/2\Phi_0$ …) in maxima of $I_c(H_{ext})$. This allows us to obtain magnetic flux dependence from external magnetic field, $\Phi(H_{ext})$. Total flux Φ is a sum of flux from external magnetic field $H_{ext}.a·L_m$ and magnetization from F-layer $4\pi M.a.L_F$ with $a$ being MJJ dimension and $L_m$ and $L_F$ effective magnetic and ferromagnetic thicknesses, respectfully. For the junction with finite thicknesses $L_1$ and $L_2$ of bottom and top electrode effective magnetic thickness $L_m$ can be approximated [55]:

$$L_m = \lambda_1 tanh(L_1/2\lambda_1) + \lambda_2 tanh(L_2/2\lambda_2) + L_F + L_{S'} + L_{Al/AlOx} \quad (2)$$

with $\lambda_1$ and $\lambda_2$ being London penetration depth for top and bottom electrodes. The inset to Fig. 6 shows magnetization curve obtained taking into account thicknesses of base, $L_1$, and top, $L_2$, electrodes; Al/AlO$_x$ layer, $L_{Al/AlOx}$; S' layer, $L_{S'}$ = 10 nm, and PdFe, $L_F$ = 14 nm, and London penetration depths for HYPRES' and ISSP processes. Adding all these thicknesses Eq. (2) gives $L_m$ = 139 nm. Magnetic moment $M$ is defined then by following:

$$4\pi M(H_{ext}) = (\Phi_0/aL_F)·(\Phi/\Phi_0) - H_{ext}·(L_m/L_F). \quad (3)$$

With $L_m$ = 139 nm, formula (3) gives magnetization dependence shown by solid circles in inset to Fig. 6. From inset to Fig. 6, the PdFe layer coercive force is ~ 12 Oe, saturation field is ~40 Oe with saturation magnetization of $4\pi M$ ~170 G.

Following the approach described in [38] $I_c(H_{ext})$ and $M(H_{ext})$ experimental data were fit by standard Fraunhofer curve. These magnetometric dependencies are presented by solid lines in Fig. 6 and inset to Fig. 6 with only $I_c(H_{ext})$ for decreasing $H_{ext}$ shown for clarity. The good agreement between experimental data and our fits proves the validity of our $L_m$ estimation. This also confirms that our SIS'FS junction acts as a single Josephson device and external magnetic field completely penetrates Al-AlO$_x$-Nb-PdFe barrier and S' layer, although being superconducting, does not screen $H_{ext}$ due to its small thickness.

## III. APPLICATION CONSIDERATIONS

As it was discussed above, MJJs can be a basis for nonvolatile random access memory, programmable digital logic arrays (functionally similar to semiconductor FPGAs), and other digital application requiring programmable change of a function. The major difference for these applications is in the array configurations. For RAM, a 2D array can be configured as a set of *serially* connected and *biased* memory cell columns. For programmable digital logic, the memory elements distributed over digital circuits. These elements can be *biased in parallel*. In both cases, addressing can be done by applying overlapping column (Y) and row (X) signals.

In the X/Y arrays, the resilience to a half-select problem is important. This is the requirement for a memory cell not to be selected when only one address (e.g., only X) is applied.

Here we are considering SIS'FS MJJs as a scalable reprogrammable element capable to energy-efficient *Read/Write* operations, long retention times, reliability, etc. Depending on S' layer thickness $L_{S'}$, there are two cases (1) of SIS'FS with $L_{S'}$ much less or comparable to $\xi_S$ and (2) with $L_{S'} >> \xi_S$.

In the first case S' layer does not participate in the screening of external magnetic field. Magnetic flux $\Phi$ in such SIS'FS junction at external magnetic field $H_{ext} = 0$ appears due to residual magnetization $M$ of F-layer. In the second case, with thick S' layer, the SIS'FS structure is very well approximated by two serially connected SIS' and S'FS junctions with $I_c$ under small exchange field, defined by SIS' $I_c$. We will consider both these SIS'FS junction types for possible applications.

### A. Read Operation

The *Read* operation is the most critical as it is typically performed more frequently that the *Write* operation. Consequently, its speed, energy efficiency, and ability for pipelining are critical. We distinguish different readout schemes applicable for RAM and programmable digital logic applications.

For RAMs, the *Read* operation is performed when MJJ is selected (i.e., biased) to have its $M(H)$ hysteresis loop shifted to have two different values of $I_c$ at $H_{ext} = 0$ [46]. The read signal then interrogates the cell and produces the voltage output depending on MJJ' $I_c$ determined by the magnetization of MJJ' F-layer. This is a simple configuration, resilient to the half-select problem. It largely reproduces the known conventional RAM schemes. As a drawback, it requires two current relatively long pulses (e.g., X-select and Y–read) overlapping in time. The generation of these pulses will be equivalent to a generation of multi-SFQ pulses. Consequently, the energy efficiency of this process will be determined by the generation of these two multi-SFQ pulses and a multi-SFQ voltage response (when "1" is read out) of the MJJ memory cell.

In order to improve the energy efficiency of the *Read* process, one needs to use lower energy single SFQ pulses (~$10^{-19}$ J). To achieve this, we can use a different scheme specific for the SFQ implementation. In this approach, only one multi-SFQ (X-select signal) can be used. The readout single SFQ pulse interrogates memory cells. A simple detection of a presence of the SFQ pulse at the output of the array slice would indicate the result of the readout. Similarly, this scheme is resilient to a half select. Its energy efficiency should be in ~ *x3* better compared to the previous scheme.

For applications in programmable logic, we can use MJJs to construct SFQ decision-making pairs (DMPs), which is the main element of RSFQ-type circuits [39]. The state of MJJs affects the DMPs thresholds to sets the desired logic function. No X/Y array readout is necessary.

### *B. Write Operation*

In contrast to the known room-temperature magnetic memory elements, our MJJs have different requirements and practical implications. The conventional MRAM technology uses the difference in resistance between two magnetized states with anti- and ferromagnetic configurations. In our case, we are using the difference in critical current of MJJs, which is maximized when one of the states is biased into an effective non-magnetized state (*M=0*) while another corresponds to a magnetized state. This is achieved when the magnetization of F-layer is combined with a magnetic self-field induced by the MJJ bias current applied for *Read* operation. This field either adds to or subtracts the F-layer magnetic field [39], [46]. We rely on magnetically soft materials, which also preferred for higher energy-efficiency remagnetization (*Write* process) as the remagnetization energy is proportional to area of hysteresis.

For both RAMs and programmable logic, the *Write* operation will be done in a similar fashion. The unbiased memory element is selected by applying one of addressed (say, X-line). Then the *Write* signal (Y line) of certain polarity (e.g., positive for "1" and negative for "0") is applied. Overlapping in time of these two pulses will add up the applied to the MJJ external magnetic field *H* and remagnetize MJJ. With presently available large multi-domain MJJ devices, the realized memory cells are not protected from a half-select problem. When the cell is half-selected, the cell will be subjected to a lower (~1/2) *Write* current causing magnetization below saturation and leaving MJJ at an undesirable magnetic state after the completion of the *Write* operation. This issue can be addressed by introduction of a magnetic anisotropy and/or reduction of MJJ size to a single-domain structure. The anisotropy will effectively transform MJJ magnetization from proportional to threshold dependence on applied *Write* current. Similarly, two magnetization directions will be summed with magnetic field induced by *Read* bias current. The scaling rules and exact dynamics of these devices will require further investigation.

The energy efficiency of this process will be determined by the generation of *X* and *Y* multi-SFQ pulses and the energy used for magnetization reversal. It is less energy-efficient than the SFQ *Read* process, which might be acceptable for typical cases with more frequent *Read* operations and in programmable logic, in which *Write* is done rarely to set the desired logic functions.

Still it would be advantageous to increase the *Write* energy efficiency and memory density. Similar to conventional magnetic memory, an injection of a spin-polarized current can be utilized instead of field bias. The memory density would improve in two ways. First, this would allow one to increase density of memory cell placement as screening layers for magnetic bias lines will not be necessary. Second, this would eliminate the need for magnetic coupling, which do not scale with the junction size reduction. We also expected that in contrast to conventional MRAM cells, the lower critical current densities for spin-polarized currents would be sufficient due to the use of soft ferromagnets. For the implementation of such spin-torque transfer memory devices, more complicated layer structure compared to SIS'FS devices discussed in this paper is required. The possible realization of these devices may include further development of recently demonstrated planar SN/FS junctions [56].

The compatibility in fabrication between SIS and SIS'FS that we demonstrated will allow us to manufacture both junction types in a single fabrication process. For successful operation within DMP, the SIS'FS critical current should be close to the SIS one. Therefore, SIS'FS junction with thicker S' layer would be preferred as their critical current is not depressed by F-layer.

Similarly, MJJs can be used as programmable bias limiting junctions to modify biases of the selected sections of ERSFQ and eSFQ circuits [5]-[7]. They also can be used in programming the SFQ clock distribution network to disable particular branches if necessary. This will lead to further reduction of dissipated power associated with SFQ clock distribution.

### IV. CONCLUSION

We investigated superconductor-insulator-superconductor-ferromagnet-superconductor (SIS'FS) Magnetic Josephson Junctions (MJJs) - the superconducting devices with ferromagnetic barrier for a scalable high-density cryogenic memory compatible with energy-efficient single flux quantum (SFQ) circuits. Specifically, we analyzed SIS'FS MJJs within framework Usadel equations and found that the properties of SIS'FS junctions fall into two distinct classes based on the thickness of S' layer. We fabricated and measured Nb-Al/AlOx-Nb-PdFe–Nb SIS'FS MJJs. These junctions are compatible with conventional SIS junctions allowing their seamless integration with SIS JJ-based ultra-fast digital SFQ circuits. Experimental results for MJJs verified their applicability for superconducting memory and digital circuits with analytical calculation for these SIS'FS structures is in good agreement with the experiment. The application of MJJ devices for memory and programmable logic circuits was discussed. Our estimates show that SIS'FS MJJs are quite suited for the use in energy-efficient RAM circuits, programmable logic arrays, and programmable clock distribution networks for energy-efficient digital circuits.

Although the MJJ based designs of different flavors known to date suggest a possibility of denser memory than that based on SQUID loops, the density will be limited by the co-location of address lines separated in-plane to avoid a magnetic cross-talk between neighboring cells. This calls for the development of spin-torque transfer approaches which would lead to the memory density defined only by memory devices. The energy efficiency of superconducting-ferromagnetic RAMs will also be strongly dependent on energy-efficiency of address decoders, address line drivers, and sense circuits. The optimization of memory cells should be done in conjunction with the optimization of these periphery circuits in order to maximize an overall energy efficient memory operation.

ACKNOWLEDGMENT

Authors are grateful to R. Hunt, J. Vivalda, D. Yohannes for the SIS trilayer wafer preparation, I. Nevirkovets and V. Semenov for useful advice, and M. Manheimer and S. Holmes for attention to this work.